\documentclass[letterpaper,10pt,conference]{ieeeconf}
\IEEEoverridecommandlockouts
\usepackage{amsmath,amssymb}
\usepackage{amsthm}
\usepackage{cite}
\usepackage{graphicx}
\usepackage{subcaption}
\usepackage{booktabs}
\usepackage{tabularx}
\usepackage{array}
\usepackage{hyperref}
\usepackage{orcidlink}
\usepackage[capitalise,noabbrev]{cleveref}
\crefname{assumption}{Assumption}{Assumptions}
\hypersetup{hidelinks}

\newcommand{\R}{\mathbb{R}}
\newcommand{\norm}[1]{\left\lVert #1 \right\rVert}

\newcommand{\trans}{^{\mathsf T}}

\newcommand{\dsafe}{d_{\mathrm{safe}}}
\newcommand{\Rc}{R_c}
\newcommand{\epsw}{\varepsilon_w}
\newcommand{\rhoH}{\rho_{\mathcal H}}

\newcommand{\ellH}{\ell_{\mathcal H}}
\newcommand{\gamH}{\gamma_{\mathcal H}}
\newcommand{\gamreq}{\gamma_{\mathrm{req}}}

\newtheorem{theorem}{Theorem}
\newtheorem{lemma}{Lemma}
\newtheorem{corollary}{Corollary}
\newtheorem{assumption}{Assumption}
\newtheorem{remark}{Remark}

\newcommand{\figplaceholder}[3]{%
  \IfFileExists{#1}{\includegraphics[width=0.98\linewidth]{#1}}{%
    \fbox{\begin{minipage}[c][#2][c]{0.94\linewidth}
      \centering\small\textbf{Figure placeholder}\\[3pt]
      \texttt{\detokenize{#1}}\\[4pt]
      #3
    \end{minipage}}}%
}

\title{Safety and Connectivity Certificates for Finite-Time Formation Transitions}

\author{Yash Purswani,
        Shivansh Agarwal
        and Anuj Tiwari
\thanks{This work has been submitted to the IEEE for possible publication.
Copyright may be transferred without notice, after which this version may
no longer be accessible.}
\thanks{The authors are with the Department of Mechanical Engineering,
Indian Institute of Technology Madras, Chennai 600036, India.
{\tt\small me22b214@smail.iitm.ac.in},
{\tt\small me24s022@smail.iitm.ac.in},
{\tt\small anujt@iitm.ac.in}}
}

\begin{document}
\maketitle

\begin{abstract}
Finite-time formation transitions are useful for coordinating multi-agent systems in time-critical missions, but convergence to a prescribed formation in finite time does not guarantee collision avoidance or communication connectivity during the transition. Existing approaches address finite-time coordination and safety-constrained formation control, yet explicit connections between disagreement during transitions, formation geometry, and coupling-gain selection need to be developed for certifying safe maneuvers. This paper develops a disagreement-based framework for certifying transient safety and connectivity during finite-time tracking of arbitrary prescribed formation trajectories. Local feedback guarantees exact finite-time tracking with a graph-independent settling-time bound, while a graph-dependent disagreement bound is used to derive sufficient geometric conditions for collision avoidance and connectivity preservation. These conditions further yield an offline coupling-gain selection rule based on the available geometric clearance. Numerical simulations illustrate the trade-off between convergence rate and control effort as the coupling gain increases. They also demonstrate the effectiveness and conservatism of the proposed certificates.
\end{abstract}


\section{Introduction}
\label{sec:intro}
Multi-agent systems enable tasks such as cooperative transportation, autonomous navigation, and distributed sensing, where multiple agents must follow a moving reference while maintaining a prescribed spatial arrangement \cite{oh2015survey,chung2018survey,fax2004information}. Formation-control frameworks commonly exploit inter-agent interactions and graph-based coordination to achieve desired
collective configurations \cite{anderson2008rigid,mesbahi2010graph}. In practical missions, however, formations may be asymmetric or time-varying, requiring  agents to track individually prescribed formation positions as the formation translates, rotates, or deforms.

For time-critical maneuvers, asymptotic convergence may be insufficient, motivating finite-time tracking with an explicit settling bound \cite{haimo1986finite,bhat2000finite}. During such maneuvers, collision avoidance and communication connectivity
must also be maintained as agents move toward their prescribed formation positions. These properties depend on the relative positions of agents, which are influenced by disagreement, i.e., differences in their individual formation tracking errors. This motivates quantifying how the disagreement evolves during finite-time transitions and relating its bounds to collision avoidance and connectivity preservation throughout the maneuver.

Existing works provide convergence guarantees for finite-time consensus and formation control  under varied interaction structures \cite{xiao2009finite,wang2010finite}. Fedele and D'Alfonso \cite{fedele2019kinematic} studied finite-time centroid tracking with bounded aggregation, while collision avoidance and connectivity preservation have been addressed through energy-based, optimization, and barrier-function approaches \cite{chen2015connectivity,filotheou2020robust,fu2022distributed,huang2024prescribed}. In contrast, this work investigates how cooperative disagreement contraction can provide explicit transient safety and connectivity certificates for individually prescribed formation trajectories, without modifying the nominal tracking protocol or requiring online safety optimization.

This paper considers a distributed first-order kinematic protocol for tracking arbitrary prescribed $C^1$ formation trajectories. Local feedback guarantees exact finite-time formation tracking with a graph-independent settling-time bound. A graph-dependent analysis shows that the relative tracking errors are bounded, which is then translated into sufficient geometric conditions for pairwise collision avoidance and preservation of a prescribed connected spanning sub-graph. These conditions subsequently lead to an offline design rule for the coupling-gain selection, based on the planned formation geometry and initial disagreement among agents. The protocol generates velocity setpoints for low-level controllers or autopilots, providing a common kinematic framework for coordinating agents with higher order and heterogeneous dynamics. The main contributions  of this can be summarized as follows,
\begin{enumerate}
    \item We propose a distributed first-order kinematic framework
    that guarantees finite-time formation tracking with a graph-independent settling-time bound for arbitrary prescribed $C^1$ trajectories.

    \item We derive a graph-dependent bound on
    disagreement that accounts for cooperative coupling
    under undirected, time-varying communication graphs.

    \item We use the disagreement bound to establish sufficient
    conditions for collision avoidance and connectivity
    preservation, and derive an offline coupling-gain
    selection rule based on the planned formation geometry
    and initial disagreement.
\end{enumerate}

\section{Background and Problem Definition}
\label{sec:model}

Consider a multi-agent system having $n\geq 2$ agents, with position state $x_i\in\mathbb{R}^d$ in $d$ dimensions, and desired position $x_{d,i}(t)$ for each agent $i$ in a given formation, where,
\begin{equation}
x_{d,i}(t)=r(t)+p_i(t),\qquad
\delta_i(t)=x_i(t)-x_{d,i}(t),
\label{eq:station}
\end{equation}
where $r(t)\in\mathbb{R}^d$ is the reference trajectory of the formation and and $p_i(t)\in\mathbb{R}^d$ is the desired offset relative to this reference. %The reference trajectory $r(t)$ coincides with the desired swarm centroid when the offsets satisfy
%$\sum_{i=1}^{n}p_i(t)=0$. {\color{blue} Sir, should I keep this line here as it is not needed till corollary 1.}
The prescribed trajectories $x_{d,i}(t)$ are assumed to be continuously differentiable, allowing the formation to translate and deform over time.

\noindent The distributed kinematic protocol for the swarm agents can be given as,
\begin{equation}
    \dot{x}_i =
    -\delta_i-\Phi_\beta(\delta_i)
    -\sum_{j=1}^{n}w_{ij}M(\delta_i-\delta_j)
    +\dot{x}_{d,i},
    \label{eq:kinematic-model}
\end{equation}
where $\dot{x}_i$ is the ideal commanded velocity, $w_{ij}\geq0$ are interaction weights, and $M\in\mathbb{R}^{d\times d}$ is symmetric positive semi-definite coupling matrix. For $0<\beta<1$, the local component-wise fractional-power feedback is
\begin{equation}
[\Phi_\beta(z)]_k
=\operatorname{sign}(z_k)|z_k|^\beta,
\qquad k=1,\ldots,d,
\label{eq:fractional-feedback}
\end{equation}
with $\Phi_\beta(0)=0$.

In a network of mobile agents, communication is usually limited
to nearby agents, and the available links may change as the agents
move. We describe the communication network by a time-varying
undirected graph $\mathcal G(t)=(\mathcal V,\mathcal E(t))$,
where $\mathcal V=\{1,\ldots,n\}$ is the set of agents and
$\mathcal E(t)$ contains the active communication links
\cite{olfati2004consensus,ren2005consensus}.
Links may be established or lost as inter-agent distances cross
the communication range, changing the graph topology while
agent state $x_i$ remains continuous.

\noindent The inter-agent interaction weights $w_{ij}$ are designed as
\begin{equation}
    w_{ij}(x)=
    \frac{a_{ij}(x)}
    {\|x_i-x_j\|^\nu+\varepsilon_w},
    \qquad \nu\geq1,\quad \varepsilon_w>0,
    \label{eq:weights}
\end{equation}
where
\begin{equation}
a_{ij}(x)=
\begin{cases}
1,&\|x_i-x_j\|\leq R_c,\\
0,&\|x_i-x_j\|>R_c,
\end{cases}
\quad a_{ij}=a_{ji},\quad a_{ii}=0.
\label{eq:range-graph}
\end{equation}
Here $R_c$ is the communication radius and $\varepsilon_w$ prevents singular weights. 

With symmetric weights $w_{ij}=w_{ji}\geq0$ and $w_{ii}=0$,
the weighted Laplacian $L_w$ is given by
\begin{equation}
    L_w=\operatorname{diag}\left(\sum_{j=1}^{n}w_{ij}\right)-W,
    \qquad W=[w_{ij}].
    \label{eq:laplacian}
\end{equation}

% From \eqref{eq:station}, the relative errors satisfy
% \begin{equation}
%     \delta_i-\delta_j
%     =(x_i-x_j)-(x_{d,i}-x_{d,j}),
%     \label{eq:distributed-error}
% \end{equation}
% allowing the coupling to be computed from relative positions and formation position assignments.

Defining the stacked error vector as
$\boldsymbol{\delta}=[\delta_1^\top,\ldots,\delta_n^\top]^\top$,
the error dynamics can be written as
\begin{equation}
    \dot{\boldsymbol{\delta}}
    =-\boldsymbol{\delta}-\Phi_\beta(\boldsymbol{\delta})
    -(L_w(x)\otimes M)\boldsymbol{\delta},
    \label{eq:error-dynamics}
\end{equation}
where $\Phi_\beta(\boldsymbol{\delta})$ denotes the stacked fractional-power feedback.

As the agents move, switching of the communication links introduces discontinuities in the interaction weights and the error dynamics. To establish the existence of solutions under network with switching topology, the dynamics are interpreted in the Filippov sense,
$\dot{\boldsymbol{\delta}}\in
\mathcal F[f](t,\boldsymbol{\delta})$,
where $f$ is the right-hand side of \eqref{eq:error-dynamics} and
\[
\mathcal F[f](t,\boldsymbol{\delta})
=
\bigcap_{r>0}
\bigcap_{N:\,|N|=0}
\overline{\operatorname{co}}
\left\{
f(t,z):z\in B(\boldsymbol{\delta},r)\setminus N
\right\}.
\]
Since the interaction weights are bounded, $f$ is measurable,
locally essentially bounded, and grows at most linearly.
Therefore, a Filippov solution exists for every finite initial
condition \cite{cortes2008discontinuous}.
The coupling term can then be expressed using an effective
Laplacian $\widetilde L_w$, which is a convex combination of
the limiting weighted Laplacians.
It follows that
$\widetilde L_w^\top=\widetilde L_w\succeq0$
and $\widetilde L_w\mathbf{1}=0$.

\begin{assumption}
\label{ass:graph}
Each agent obtains its own state $x_i$ through onboard odometry
and has access to its prescribed formation trajectory $x_{d,i}$
and its derivative $\dot{x}_{d,i}$.
The local tracking error $\delta_i$ can therefore
be computed onboard. Each agent shares its tracking error
$\delta_i$ with its active neighbours through local communication,
allowing it to compute the relative errors $\delta_i-\delta_j$.
\end{assumption}

\noindent\textbf{Problem statement:}
The research problem addressed in this paper is to develop a distributed kinematic protocol that 

\noindent
(i) achieves exact finite-time formation position tracking with a graph-independent settling-time bound; 

\noindent
(ii) establishes sufficient conditions for pairwise collision avoidance and preservation of the connected spanning sub-graph $\mathcal H$; and 

\noindent
(iii) provides an offline coupling-gain selection from the planned desired formation geometry and initial disagreement.

\section{Finite-Time Formation Tracking}
\label{sec:finite-time}
This section establishes finite-time tracking of the prescribed formation trajectories under the proposed kinematic protocol. Explicit bounds on the tracking error and 
settling-time are derived along with finite-time centroid tracking.

Let $\delta_0=\boldsymbol{\delta}(t_0)$ denote the initial
formation tracking error vector. The error bound is
defined as
\begin{equation}
\begin{aligned}
\mathcal R(t)&=\left[\bigl(1+\norm{\delta_0}^{1-\beta}\bigr)
 e^{-(1-\beta)(t-t_0)}-1\right]^{1/(1-\beta)}_+,
\end{aligned}
\label{eq:R-bound}
\end{equation}
where $[a]_+=\max\{a,0\}.$
The finite-time arrival bound is
\begin{equation}
  T=t_0+\frac{\ln(1+\norm{\delta_0}^{1-\beta})}{1-\beta}.
  \label{eq:T-bound}
\end{equation}

\begin{theorem}[Finite-time formation tracking]
\label{thm:station}
Under \cref{ass:graph}, with $M=M\trans\succeq0$ and $0<\beta<1$, every Filippov solution of \eqref{eq:error-dynamics} satisfies
\begin{equation}
  \norm{\boldsymbol{\delta}(t)}\leq\mathcal R(t),\qquad t \geq t_0
  \label{eq:station-envelope}
\end{equation}
and $\boldsymbol{\delta}(t)=0$ for all $t\geq T$. Thus, $x_i(t)=x_{d,i}(t)$ for all agents and all $t\geq T$. The bound is independent of the interaction graph, its connectivity, and $M$.
\end{theorem}

\begin{proof}
Consider $V=\frac12\norm{\boldsymbol{\delta}}^2$. For almost every time, symmetry and positive semi-definiteness of the effective Laplacian and $M$ give
\begin{equation}
\begin{aligned}
  \mathcal D&:=\boldsymbol{\delta}\trans(\widetilde L_w\otimes M)\boldsymbol{\delta}\\
   &=\frac12\sum_{i,j}\widetilde w_{ij}(\delta_i-\delta_j)\trans M
      (\delta_i-\delta_j)\geq0.
\end{aligned}
\label{eq:edge-energy}
\end{equation}
Since $1+\beta<2$, norm monotonicity implies
$\sum_{\ell=1}^{nd}|\delta_\ell|^{1+\beta}\ge\norm{\boldsymbol{\delta}}_2^{1+\beta}$.
Consequently along \eqref{eq:error-dynamics},
\begin{align}
  \dot V&=-\norm{\boldsymbol{\delta}}^2-\boldsymbol{\delta}\trans\Phi_\beta(\boldsymbol{\delta})-\boldsymbol{\delta}\trans(\widetilde L_w\otimes M)\boldsymbol{\delta}\\
  &\leq-\norm{\boldsymbol{\delta}}^2-\norm{\boldsymbol{\delta}}^{1+\beta}
   =-2V-(2V)^{(1+\beta)/2},
  \label{eq:V-station}
\end{align}
Let $z=\norm{\boldsymbol{\delta}}=\sqrt{2V}$. Whenever $z>0$, \eqref{eq:V-station} implies $\dot z\leq-z-z^\beta$. With $y=z^{1-\beta}$,
\begin{equation}
  \dot y\leq-(1-\beta)(y+1).
\end{equation}
Integration yields $y(t)+1\leq(y(t_0)+1)e^{-(1-\beta)(t-t_0)}$ until the tracking error reaches zero. This comparison bound establishes the bounds in \eqref{eq:R-bound} and \eqref{eq:T-bound}, guaranteeing that the error reaches zero within a finite time. Since the origin is invariant under the error dynamics, the error remains zero after reaching it. Since $V$ is non-increasing, the error remains bounded.
\end{proof}

\subsection{Centroid and Formation Guarantees}
Assume the formation position offsets are centered, i.e.,
\begin{equation}
  \sum_{i=1}^{n}p_i(t)=0,\qquad \forall t\geq t_0.
  \label{eq:centered}
\end{equation}
Let $c(t)=\frac1n\sum_i x_i(t)$ be the swarm centroid. Since $x_{d,i}=r+p_i$, the mean desired formation position is $r$.

\begin{corollary}[Reference tracking]
\label{cor:reference}
Under the centering condition \eqref{eq:centered},
\begin{equation}
    \norm{c(t)-r(t)}
    \leq \frac{\mathcal R(t)}{\sqrt n},
    \qquad t\geq t_0.
\end{equation}
Consequently, $c(t)=r(t)$ for all $t\geq T$.
\end{corollary}

\begin{proof}
Since $x_{d,i}=r+p_i$ and $\sum_{i=1}^n p_i=0$, the centroid error satisfies
\[
c-r=\frac{1}{n}\sum_{i=1}^n(x_i-x_{d,i})
=\frac{1}{n}\sum_{i=1}^n\delta_i.
\]
By the Cauchy--Schwarz inequality,
\[
\|c-r\|
\leq \frac{1}{\sqrt n}
\left(\sum_{i=1}^n\|\delta_i\|^2\right)^{1/2}
=\frac{\|\boldsymbol{\delta}\|}{\sqrt n}.
\]
Applying the bound $\|\boldsymbol{\delta}(t)\|\leq\mathcal R(t)$ from Theorem~\ref{thm:station} gives the stated inequality. Since $\mathcal R(t)=0$ for all $t\geq T$, it follows that $c(t)=r(t)$ for all $t\geq T$.
\end{proof}

\begin{remark}
Note that when the prescribed offsets are not centered, the swarm centroid converges in finite time to $r(t)+\frac{1}{n}\sum_{i=1}^{n}p_i(t)$, rather than $r(t)$.
\end{remark}

\section{Disagreement-Based Safety and Connectivity Certificates}
\label{sec:certificates}
To study collision avoidance and connectivity preservation,
the effect of formation tracking errors on inter-agent distances
is considered. Since relative distances depend only on differences
between individual tracking errors, a bound on disagreement
is first derived for time-varying communication graphs.
This bound is then used to establish sufficient conditions
for collision avoidance and connectivity preservation.

\noindent The disagreement vector is defined as
\begin{equation}
\begin{aligned}
  \Pi&=I_n-\frac1n\mathbf 1\mathbf 1\trans,\qquad
  \boldsymbol{\eta}=(\Pi\otimes I_d)\boldsymbol{\delta},\\
  \mu&=\lambda_{\min}(M)\geq0.
\end{aligned}
\label{eq:disagreement}
\end{equation}
Thus $\eta_i=\delta_i-\bar\delta$ and $\delta_i-\delta_j=\eta_i-\eta_j$.

\begin{theorem}[Graph-dependent disagreement contraction]
\label{thm:disagreement}
For every Filippov solution at $t\in[t_0,T]$, the disagreement satisfies
\begin{equation}
  \norm{\boldsymbol{\eta}(t)}
  \leq \norm{\boldsymbol{\eta}(t_0)}
  \exp\left[
    -\int_{t_0}^{t}
    \left(1+\mu\lambda_2(\widetilde L_w(s))\right)\,ds
  \right].
  \label{eq:eta-bound}
\end{equation}
Moreover, $\boldsymbol{\eta}(t)=0$ for all $t\geq T$. The bound remains valid across communication topology changes.
\end{theorem}

\begin{proof}
Since $\Pi\widetilde L_w=\widetilde L_w$ and $(\Pi\otimes I_d)\boldsymbol{\eta}=\boldsymbol{\eta}$,
the disagreement dynamics follow from \eqref{eq:error-dynamics} as
\begin{equation}
  \dot{\boldsymbol{\eta}}=-\boldsymbol{\eta}-(\Pi\otimes I_d)\Phi_\beta(\boldsymbol{\delta})
  -(\widetilde L_w\otimes M)\boldsymbol{\eta}.
\end{equation}
For $W_\eta=\frac12\norm{\boldsymbol{\eta}}^2$, using
$\boldsymbol{\eta}\trans(\Pi\otimes I_d)=\boldsymbol{\eta}\trans$ yields
\begin{equation}
  \dot W_\eta=-\norm{\boldsymbol{\eta}}^2-\boldsymbol{\eta}\trans\Phi_\beta(\boldsymbol{\delta})
  -\boldsymbol{\eta}\trans(\widetilde L_w\otimes M)\boldsymbol{\eta}.
\end{equation}
By component-wise monotonicity of $\Phi_\beta$,
\begin{equation}
\begin{aligned}
  \boldsymbol{\eta}\trans\Phi_\beta(\boldsymbol{\delta})
  &=\frac{1}{2n}\sum_{i,j}
  (\delta_i-\delta_j)^\top
  \big(\Phi_\beta(\delta_i)-\Phi_\beta(\delta_j)\big)
  \geq 0.
\end{aligned}
\end{equation}
Since $M\succeq0$ and $\mu=\lambda_{\min}(M)$,
the Rayleigh quotient bound gives
\[
  \boldsymbol{\eta}^\top(\widetilde L_w\otimes M)\boldsymbol{\eta}
  \geq \mu\lambda_2(\widetilde L_w)\norm{\boldsymbol{\eta}}^2.
\]
Consequently,
\[
  \dot W_\eta
  \leq -2\bigl[1+\mu\lambda_2(\widetilde L_w)\bigr]W_\eta.
\]
Applying the comparison lemma and integrating yields
\eqref{eq:eta-bound}. 
% Since the inequality holds almost
% everywhere along Filippov solutions, the bound remains valid across topology changes.
Finally, $\Pi$ is an orthogonal
projection, so $\norm{\boldsymbol{\eta}(t)}\leq\norm{\boldsymbol{\delta}(t)}$.
By Theorem~\ref{thm:station}, $\boldsymbol{\delta}(t)=0$ for all $t\geq T$,
which implies $\boldsymbol{\eta}(t)=0$ for all $t\geq T$.
\end{proof}

\subsection{Pairwise Distance Bounds}
\noindent Let $d_{ij}^q(t)=\norm{x_{d,i}(t)-x_{d,j}(t)}$ and $d_{ij}^x(t)=\norm{x_i(t)-x_j(t)}$.

\begin{lemma}[Relative-distance envelope]
\label{lem:distance-tube}
For every pair $i,j$,
\begin{equation}
  d_{ij}^q(t)-\sqrt2\norm{\boldsymbol{\eta}(t)}\leq d_{ij}^x(t)
  \leq d_{ij}^q(t)+\sqrt2\norm{\boldsymbol{\eta}(t)}.
  \label{eq:distance-tube}
\end{equation}
\end{lemma}
\begin{proof}
Since $x_i-x_j=(x_{d,i}-x_{d,j})+(\eta_i-\eta_j)$, the triangle inequality gives
\[
\|x_i-x_j\|
\leq \|x_{d,i}-x_{d,j}\|+\|\eta_i-\eta_j\|,
\]
while the reverse-triangle inequality yields
\[
\|x_i-x_j\|
\geq \|x_{d,i}-x_{d,j}\|-\|\eta_i-\eta_j\|.
\]
Moreover,
\[
\begin{aligned}
\norm{\eta_i-\eta_j} \leq \norm{\eta_i}+\norm{\eta_j} \leq\sqrt{2(\norm{\eta_i}^2+\norm{\eta_j}^2)}\leq\sqrt2\norm{\boldsymbol{\eta}}. 
\end{aligned}
\]
Hence, the two inequalities above establish the claimed bounds.
\end{proof}

A connected spanning sub-graph $\mathcal H=(\mathcal V,\mathcal E_{\mathcal H})$ is selected for preservation to maintain communication connectivity throughout formation maneuvers. Let $L_{\mathcal H}$ denote its unweighted Laplacian. 
Define the maximum prescribed edge length over $\mathcal{H}$ and the minimum prescribed pairwise distance as
\begin{equation}
\begin{aligned}
  \ellH(t)&=\max_{(i,j)\in\mathcal E_\mathcal H}\norm{x_{d,i}(t)-x_{d,j}(t)},\\
  d_q(t)&=\min_{i<j}\norm{x_{d,i}(t)-x_{d,j}(t)}.
\end{aligned}
\label{eq:station-geometry}
\end{equation}
Let $\lambda_2(L_{\mathcal H})>0$ and let $\underline w>0$ be a lower bound on every active selected sub-graph-edge weight. For the range-dependent weights in \eqref{eq:weights}, one may choose $\underline w=(\Rc^\nu+\epsw)^{-1}$. Define
\begin{equation}
\begin{aligned}
  \gamH&=1+\mu\underline w\lambda_2(L_{\mathcal H}),\\
  \rho_0&=\sqrt2\norm{\boldsymbol{\eta}(t_0)}.
\end{aligned}
\label{eq:gain-definitions}
\end{equation}
Since $\boldsymbol{\eta}=(\Pi\otimes I_d)\boldsymbol{\delta}$ and $\Pi$ is an orthogonal projection, it cannot increase the Euclidean norm; hence $\|\boldsymbol{\eta}(t)\|\leq\|\boldsymbol{\delta}(t)\|$. By \cref{thm:station}, this gives $\sqrt2\|\boldsymbol{\eta}(t)\|\leq\sqrt2\,\mathcal R(t)$. Moreover, \cref{thm:disagreement} gives exponential contraction of the disagreement. While all selected sub-graph edges remain active, $\widetilde L_w\succeq\underline w L_{\mathcal H}$, so the contraction rate is at least $\gamH$. Consequently,
\begin{equation}
  \rhoH(t)=\min\left\{\sqrt2\,\mathcal R(t),\,
  \rho_0e^{-\gamH(t-t_0)}\right\}.
  \label{eq:rho-envelope}
\end{equation}
Thus $\rhoH(t)$ provides an a priori bound on the maximum pairwise disagreement for every pair of agents, provided the selected sub-graph remains available.

\begin{theorem}[Joint collision avoidance and connectivity preservation]
\label{thm:safety-connectivity}
If the planned formation positions satisfy, for  all $t\geq t_0$,
\begin{align}
  d_q(t)&>\dsafe+\rhoH(t), \label{eq:safe-condition}\\
  \ellH(t)&<\Rc-\rhoH(t), \label{eq:connect-condition}
\end{align}
then every pair of agents remains strictly separated by more than $\dsafe$, and all edges of the prescribed spanning sub-graph $\mathcal{H}$ remain within communication range $\Rc$. 
% The formation position-tracking bound $T$ in \eqref{eq:T-bound} is unchanged.
\end{theorem}

\begin{proof}
At $t_0$, \eqref{eq:rho-envelope} gives $\rhoH(t_0)=\rho_0$, since $\norm{\boldsymbol{\eta}(t_0)}\leq\norm{\boldsymbol{\delta}(t_0)}$. By \cref{lem:distance-tube} and \eqref{eq:safe-condition}, all actual pairwise distances exceed $\dsafe$ initially. The same lemma and \eqref{eq:connect-condition} put all selected sub-graph links strictly inside $\Rc$ initially.

Suppose a selected sub-graph edge first reaches $\Rc$ at time $t_e>t_0$. Until $t_e$, all selected sub-graph edges remain active, so the lower bound $\widetilde L_w\succeq\underline wL_{\mathcal H}$ and \cref{thm:disagreement} imply \eqref{eq:rho-envelope}. By continuity, the bound also holds at $t_e$. For every selected sub-graph edge, \cref{lem:distance-tube} then gives $\norm{x_i(t_e)-x_j(t_e)}\leq\ellH(t_e)+\rhoH(t_e)<\Rc$, a contradiction. Thus the selected sub-graph persists for all time. Finally, the lower half of \cref{lem:distance-tube} and \eqref{eq:safe-condition} imply $d_{ij}^x(t)>\dsafe$ for every pair and all times. The formation position-tracking theorem is unaffected by the graph argument.\end{proof}

\subsection{Coupling-Gain Design Rule}
Define the available geometric margin
\begin{equation}
  s(t)=\min\{d_q(t)-\dsafe,\,\Rc-\ellH(t)\}.
  \label{eq:margin}
\end{equation}
The margin depends only on the prescribed formation position trajectories, the selected safety threshold, the operational communication range, and the selected sub-graph. It can therefore be computed offline, before execution, without live agent-state measurements. Geometric feasibility requires $s(t)>0$ for all $t\geq t_0$.
Suppose $0<\rho_0<s(t_0)$. Since exact formation tracking is achieved by $T$, the gain need only compensate for disagreement before $T$. Define
\begin{equation}
  \gamreq=\sup_{t_0<t<T}\frac{\ln\big(\rho_0/s(t)\big)}{t-t_0}.
  \label{eq:required-rate}
\end{equation}

\begin{corollary}[Sufficient coupling-gain selection]
\label{cor:gain}
A sufficient condition for \cref{thm:safety-connectivity} is
\begin{equation}
  1+\lambda_{\min}(M)\,\underline w\,\lambda_2(L_{\mathcal H})>\gamreq.
  \label{eq:gain-condition}
\end{equation}
For $M=mI_d$, if $\gamreq>1$, it is sufficient to choose
\begin{equation}
 m>\frac{\gamreq-1}{\underline w\lambda_2(L_{\mathcal H})}.
 \label{eq:gain-explicit}
\end{equation}
Thus, once the formation position plan and physical limits $(d_{\mathrm{safe}},R_c)$ are specified, the margin and required rate can be computed offline and used to select a gain satisfying the certificate.
\end{corollary}

\begin{proof}
By \cref{thm:safety-connectivity}, it suffices to ensure that
$\rhoH(t)<s(t)$ throughout the maneuver. By the definition of
$\gamma_{\mathrm{req}}$ in \eqref{eq:required-rate},
$\gamH>\gamma_{\mathrm{req}}$ ensures that
\[
\rho_0e^{-\gamH(t-t_0)}<s(t),
\qquad t\in(t_0,T).
\]
Since $\rhoH(t)\leq\rho_0e^{-\gamH(t-t_0)}$ by
\eqref{eq:rho-envelope}, the sufficient conditions hold on
$(t_0,T)$. The strict initial condition handles $t=t_0$.
For $t\geq T$, exact formation tracking gives $\rho_{\mathcal H}(t)=0<s(t)$ by geometric feasibility.
Thus, the conditions of \cref{thm:safety-connectivity}
hold for all $t\geq t_0$. In particular, its minimum
margin
\begin{equation}
\Delta_{\min}:=\min_{t\in[t_0,T]}
\left[s(t)-\rhoH(t)\right]
\end{equation}
is strictly positive, quantifying the available clearance
of the certificate.
\end{proof}

\begin{figure*}[t]
\centering
\includegraphics[width=0.9\linewidth]{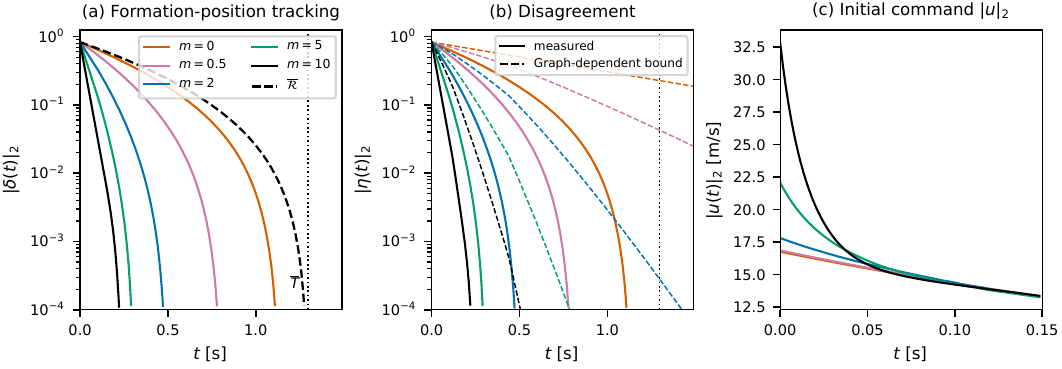}
\caption{Finite-time tracking, disagreement decay, and commanded-velocity magnitude across coupling gains $m$ ($\kappa=2.0$): (a) $\|\boldsymbol{\delta}(t)\|_2$ and theoretical bound $\overline{\mathcal R}(t)$ with settling-time bound $T$; (b) $\|\boldsymbol{\eta}(t)\|_2$ and graph-dependent  bounds; (c) aggregate commanded-velocity magnitude $\|u(t)\|_2$ over the initial $0.15$~s.}
\label{fig:convergence}
\end{figure*}

\begin{remark}
Note that when $\gamma_{\mathrm{req}}<1$, the conditions can be satisfied with $M=0$, provided geometric feasibility and the initial-margin condition hold. The gain condition is sufficient but not necessary, and cannot compensate for a violation of the initial-margin condition $\rho_0<s(t_0)$.
\end{remark}

\section{Numerical Validation}
\label{sec:numerics}
Following the simulation setup and parameters in
\cref{tab:sim-parameters}, two sets of numerical experiments
are conducted with $n=10$ agents in $\mathbb{R}^2$:
(i) finite-time formation tracking and the effect of coupling
gain on convergence and control effort, and
(ii) collision avoidance and connectivity preservation during
rapid formation compression and expansion.

Both experiments consider a rotating star formation with
angular velocity $\omega$ and time-varying scaling
$a(t)=a_\infty+(a_0-a_\infty)e^{-\kappa t}$ of the formation.
The initial perturbations satisfy $\rho_0<s(0)$.
For connectivity analysis, a connected spanning sub-graph
with 15 edges and $\lambda_2(L_{\mathcal H})\approx0.9316$
is selected.
The simulation framework and experiments are available at this
{\color{blue}\underline{\href{https://github.com/yash-purswani/finite-time-swarm-control}{github repository}}}.

\begin{table}[t]
\centering
\caption{Benchmark simulation parameters.}
\label{tab:sim-parameters}
\footnotesize
\begin{tabularx}{\columnwidth}{llX}
\toprule
Param. & Value & Description \\
\midrule
$n$ & $10$ & Swarm size in $\R^2$ \\
$\beta$ & $0.5$ & Fractional-power velocity exponent \\
$\dsafe$ & $1.0$~m & Minimum pairwise collision threshold \\
$\Rc$ & $6.0$~m & Maximum communication radius \\
$\epsw$ & $0.05$ & Proximity weight regularizer \\
$\nu$ & $1$ & Distance exponent in interaction weights \\
$\underline{w}$ & $0.1653$~m$^{-1}$ & Uniform lower bound $1/(\Rc + \epsw)$ \\
$r(t)$ & $[t,\ 2\sin(0.5t)]\trans$ & Reference centroid trajectory \\
$\omega$ & $\pi/4$~rad/s & Star template angular rotation rate \\
$M$ & $m I_2$ & Cooperative coupling matrix ($m \ge 0$) \\
$\mathcal H$ & 15 edges & Five-cycle on inner agents; each outer joins two adjacent inner agents \\
Solver & DOP853 & $\texttt{rtol}=10^{-6}$, $\texttt{atol}=10^{-8}$, $\texttt{max\_step}=0.02$~s \\
\bottomrule
\end{tabularx}
\end{table}

\subsection{Finite-Time Tracking and Coupling-Gain Trade-offs}

The effect of coupling gain $m$ is evaluated during a rotating
star compression ($\kappa=2.0$, $a_0=1.8$, $a_\infty=0.8$).
As shown in \cref{fig:convergence,tab:gain-sweep}, the tracking
error converges within the theoretical settling-time bound
$T=1.3621$~s for all tested gains.
The numerical settling time $t_{10^{-6}}$ is measured when
$\|\boldsymbol{\delta}(t)\|_2$ first falls below $10^{-6}$.
Increasing $m$ leads to faster tracking and disagreement
convergence, while the theoretical settling-time bound
remains unchanged.

To examine the associated control effort, the commanded
velocity is defined as $u_i=\dot{x}_i$, with aggregate magnitude
$\|u(t)\|_2=(\sum_i\|u_i(t)\|^2)^{1/2}$.
The peak commanded velocity and the integrated squared
velocity $J_u=\int_{t_0}^{t_f}\|u(t)\|_2^2\,dt$
are used as measures of control effort.
As shown in \cref{tab:gain-sweep}, higher coupling gains
reduce the numerical settling time but increase the
commanded-velocity magnitude and overall control effort.

\begin{table}[t]
\centering
\caption{Settling times and aggregate commanded-velocity metrics across coupling gains.}
\label{tab:gain-sweep}
\small
\begin{tabular}{ccccc}
\toprule
$m$ & $t_{10^{-6}}$ (s) & $T$ (s) & Peak $\norm{u}_2$ [m/s] & $J_u$ [m$^2$/s]\\
\midrule
 0.0 & 1.1517 & 1.3621 & 16.9828 & 429.44 \\
 0.5 & 0.8117 & 1.3621 & 17.0860 & 429.92 \\
 2.0 & 0.5250 & 1.3621 & 18.0322 & 431.55 \\
 5.0 & 0.4467 & 1.3621 & 22.2186 & 434.95 \\
10.0 & 0.3933 & 1.3621 & 32.8156 & 440.74 \\
\bottomrule
\end{tabular}
\end{table}

\subsection{Transient Safety and Connectivity}
To evaluate the proposed offline gain-selection rule, we consider aggressive star compression ($a_0=1.80$ to $a_\infty=0.80$) and expansion ($a_\infty=3.00$) at $\kappa=15.0$. As shown in Fig.~2 and Table~III, the selected gain $m=70$ satisfies the numerically evaluated safety and connectivity certificates in both maneuvers, whereas weak coupling ($m=1$) results in collision and connectivity violations, respectively. Notably, intermediate coupling ($m=30$) preserves both constraints despite failing the sufficient certificate, illustrating its conservatism. 
The reported $\Delta_{\min}$ values are estimated from sampled trajectories. Video demonstrations of the formation compression and expansion simulation experiments are available at this
{\color{blue}\underline{\href{https://drive.google.com/drive/folders/1a-hdQOMJ1HW1a-QlcL2H3Y5TU_pJnpCx?usp=sharing}{link}}}.

\begin{figure*}[t]
    \centering
    \begin{subfigure}[t]{0.48\textwidth}
        \centering
        \includegraphics[width=\linewidth]{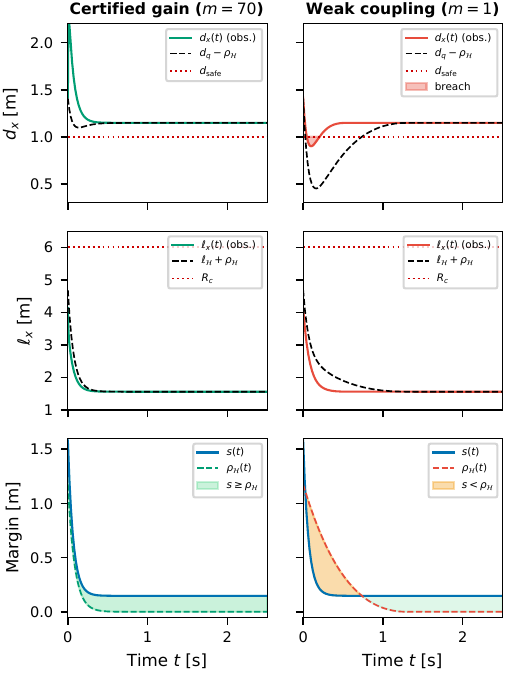}
        \caption{Formation compression ($a_0 = 1.8 \to a_\infty = 0.8$).}
        \label{fig:safety-transient-compression}
    \end{subfigure}
    \hfill
    \begin{subfigure}[t]{0.48\textwidth}
        \centering
        \includegraphics[width=\linewidth]{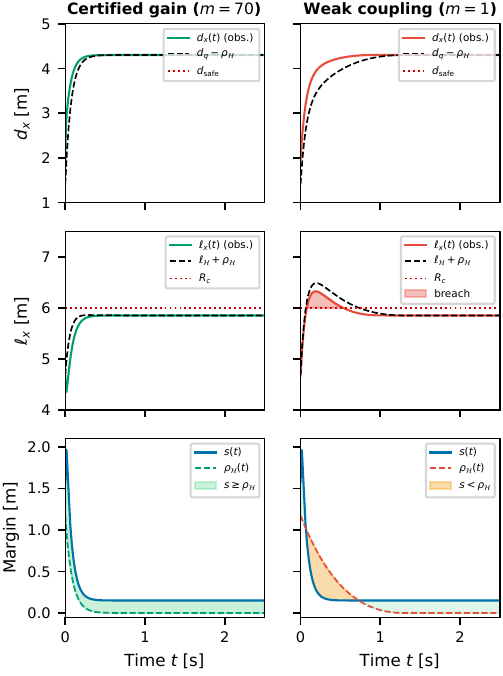}
        \caption{Formation expansion ($a_0 = 1.8 \to a_\infty = 3.0$).}
        \label{fig:safety-transient-expansion}
    \end{subfigure}

    \caption{Transient safety and connectivity certificates during star compression and expansion ($\kappa=15.0$). Columns compare certified ($m=70$) and weak ($m=1$) coupling; rows show the minimum pairwise distance, maximum selected-edge distance, and the geometric clearance $s(t)$ against the disagreement envelope $\rhoH(t)$. The certificate holds when $s(t)>\rhoH(t)$. Weak coupling leads to a collision during compression and a communication-range violation during expansion.}
    \label{fig:transient-safety-connectivity}
\end{figure*}

\begin{table*}[t]
\centering
\caption{Numerically evaluated certificates and observed geometry for aggressive star maneuvers ($\kappa=15.0$).}
\label{tab:safety-comparison}
\resizebox{\textwidth}{!}{%
\begin{tabular}{llrrrrrrrrl}
\toprule
Maneuver & Case & $m$ & $\kappa$ & $\gamH$ & $\gamreq$ & $t_{10^{-6}}$ (s) & $d_{x,\min}$ (m) & $\ell_{x,\max}$ (m) & $\Delta_{\min}$ (m) & Observed outcome\\
\midrule
Compression & Certified gain & 70 & 15 & 11.7786 & 9.3136 & 0.1750 & 1.1473 & 4.1387 & \textbf{0.0970} & Certified; safe, connected \\
            & Uncertified gain & 30 & 15 & 5.6194 & 9.3136 & 0.1783 & 1.1473 & 4.1387 & -0.2150 & Observed safe, connected \\
            & Weak coupling & 1 & 15 & 1.1540 & 9.3136 & 0.5567 & 0.8990 & 4.1387 & -0.5500 & Collision observed \\
\midrule
Expansion & Certified gain & 70 & 15 & 11.7786 & 7.4220 & 0.2300 & 1.9785 & 5.8490 & \textbf{0.1427} & Certified; safe, connected \\
            & Uncertified gain & 30 & 15 & 5.6194 & 7.4220 & 0.4000 & 1.9785 & 5.8490 & -0.1180 & Safe, Observed connected \\
            & Weak coupling & 1 & 15 & 1.1540 & 7.4220 & 1.0450 & 1.9785 & 6.3214 & -0.4836 & Range violation observed \\
\bottomrule
\end{tabular}}
\end{table*}

\section{Conclusion}
This paper developed a framework for finite-time tracking of arbitrary prescribed formation trajectories, with transient safety and connectivity certificates. Local feedback guarantees exact formation tracking within a graph-independent settling-time bound, while cooperative coupling accelerates disagreement decay without altering this guarantee. The disagreement bound yields sufficient conditions for collision avoidance and preservation of a selected connected spanning sub-graph, together with an offline coupling-gain selection rule based on planned geometric clearance and initial disagreement. Numerical experiments validate the common finite-time settling bound across coupling gains and illustrate the trade-off between faster convergence and increased control effort. Rapid formation compression and expansion illustrate how appropriately selected coupling gains can certify collision avoidance and connectivity, while insufficient coupling may lead to constraint violations. The certificates are sufficient but not necessary, and their failure does not imply unsafe behaviour. Although the gain-selection rule provides a lower bound on coupling, actuator limits may impose practical upper bounds. Extending these guarantees to higher-order dynamics and actuator-constrained systems remains future work.

\bibliographystyle{IEEEtran}
\bibliography{references}
\end{document}